\documentclass[12pt,preprint]{aastex}

\begin{document}

\shorttitle{Astrophysical Jets and Heterogeneous Media}
\shortauthors{Yirak et al.}

\title{The Interaction Between a Pulsed Astrophysical Jet and Small-Scale Heterogeneous Media}
\author{Kristopher Yirak\altaffilmark{1}, Adam Frank, Andrew Cunningham}
\affil{Department of Physics and Astronomy, University of Rochester,
    Rochester, NY 14620}
\and
\author{Sorin Mitran}
\affil{Department of Mathematics, Applied Mathematics Program, University of North Carolina, Chapel Hill, NC 27599}
\altaffiltext{1}{Email contact: yirak@pas.rochester.edu}

\begin{abstract}
We have performed 2D hydrodynamic simulations of a pulsed astrophysical jet propagating through a medium that is populated with spherical inhomogeneities, or ``clumps,'' which are smaller than the jet radius. The clumps are seen to affect the jet in several ways, such as impeding jet propagation and deflecting the jet off-axis. While there has been some debate as to the prevalence of these types of condensations in the ISM or in molecular clouds, the exploration of this region of parameter space nonetheless both shows the potential for these clumps to disrupt astrophysical jets and yields results which recover aspects of recent observations of Herbig-Haro objects. We find that the propagation of the jet and the vorticity induced in the clump/ambient medium correlate well with a ``dynamic filling function'' $f_d$ across all the simulations.
\end{abstract}


\section{Introduction}
The investigation of astrophysical flows, such as those of Herbig-Haro objects, AGN jets, and planetary nebulae has benefited in recent years from the advancement of computing capability.  Hydrodynamic and magnetohydrodynamic simulations of HH objects have been performed in 2D \citep[e.g.][]{raga2002,osullivan2002} and 3D \citep[e.g.][]{dalpino1999} for a variety of parameters.  The effect of heterogeneous media on astrophysical flows has also received attention in terms of simulation studies.  Adiabatic and non-adiabatic simulations of astrophysically-relevant shocked clouds have been performed by \cite{poludnenko2002} and \cite{fragile2004}, respectively, as have simulations of extragalactic radio jets in ``clumpy'' media \citep{saxton2005}.  Planetary nebulae simulations have also been carried forward which include the effect of clumps \citep{steffen2004}. Typical motivations in these simulations are both the recovery of specific morphology and the capacity to characterize global characteristics, such as detailed velocity distributions.

Observations of HH objects show them to be highly-collimated, several-hundred-AU radius structures with a primary bowshock 0.3 pc to several pc from their source with subsequent bowshocks or ``knot'' features trailing behind (see e.g.~\cite{hartigan2005} or the review by \cite{reipurth2001}).  Detailed images of the bowshock reveal complex, dynamic structure. By comparing observations made 3 years apart, \cite{bally2002} showed that the bowshock of HH 1 displays a wide range of motion in addition to its bulk flow.  Using the two datasets, they were able to derive velocity vectors for the bowshock; see Fig. 1. The observations of multiple clumps in the flow made them particularly compelling though it was not clear if the clumps were fragments of the jet/bow shock disrupted via instabilities or environmental density perturbations enhanced by the passage of the shock.

\begin{figure}[htbp]
\epsscale{0.5}
\plotone{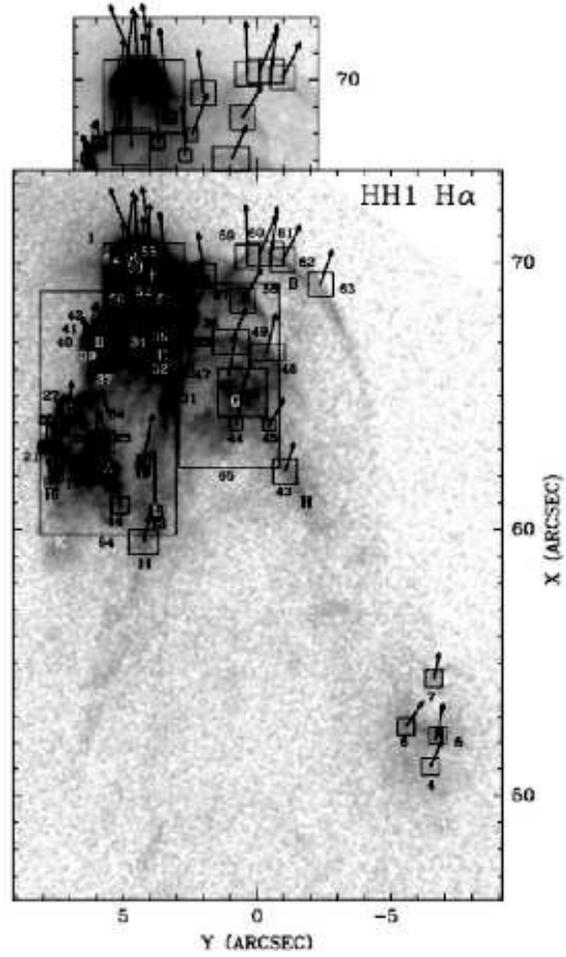}
\caption{1994 epoch $HST$ image of the HH 1 bowshock. The length of arrows corresponds to motion over 15 years. From \cite{bally2002}.}
%\label{f1}
\end{figure}

Of particular concern with HH objects is the relationship between the jet morphology and its environment.  Jets which show large deflections have been argued to evolve from a precessing \citep{masciadri2002} or otherwise variable source \citep{cabrit2000}. \cite{rosen2004} performed pulsed jet simulations in which the density ratio between jet and ambient was varied. Earlier, \cite{volker1999} performed a study in which pulsation, shear, precession, and opening angle were all varied. Further discussion of variable source studies may be found therein.

While a variable source may indeed account for some ``bending'' of jets, there are other cases where the morphology demands an alternate explanation, such as HH 110.  These explanations have typically included a significant interaction with an object in the jet's path \citep{dalpino1999,raga2002} or with a strong ambient magnetic field \citep{decolle2005}.

Thus the role of clumps influencing HH jets and HH objects is of interest for both observational and purely theoretical reasons.  Jet interactions with heterogeneous media in other contexts such as planetary nebula are of interest as well for similar reasons.  To date, simulations involving HH jets and clumps  have considered inhomogeneities which are larger than the jet itself \citep[see][]{raga1987,dalpino1999}.  In this paper, we present results from 2D hydrodynamic simulations involving dense ``clumps'' which are 30 AU in radius---a fraction of the jet's own 100 AU radius.

The existence of dense condensations on the order of 10--100 AU has been the subject of some debate. In the ISM, \cite{heiles1997} asserted that features which appeared to be these small clumps could in fact be larger structures---filaments or sheets---which are viewed edge-on.  \cite{deshpande2000} suggested that the supposed observations of clumps of this size could in fact be accounted for by a power law extension of irregularities in opacities on larger scales in the ISM.  However, \cite{faison2001} presented results using very long baseline interferometry observations of several extragalactic sources which they purported to show several distinct clumps less than 100 AU in extent.  More recently, \cite{hennebelle2006} presented work arguing that the presence of Alfv\'en waves may serve to enhance the formation of small-scale structure in the ISM.

In star-forming regions, detection limits play an increased role. \cite{vannier2001} observed H$_2$ from a region of the Orion molecular cloud with a resolution of $\sim$0.15'', corresponding to physical length of $\sim$50 AU. However, synthetic data used for calibration of Fourier analysis created spurious signals below $\sim$100 AU, and so data below this level were not used. With regards to the prevalence of small-scale structure in star-forming regions, the situation seems even less clear. The survey of \cite{hatchell2005} of the Perseus molecular cloud reported beam mass sensitivity at 12 K only down to 0.4M$_\odot$---the structures here in question are considerably smaller.

At the very least, the observation and explanation of media on this small scale has proven to be difficult.  While the existence of clumps on this scale remains an open question, the investigation in this area of the parameter regime remains of interest.  As we shall see the propagation of radiative jet in a clumpy media yields some new results which will be of general interest.

\section{Computational Method \& Physical Model}
We here employ the 2D (slab-symmetric) hydrodynamic, radiatively-cooling AMR capability of the \emph{AstroBEAR} computational code\footnote{Information about the BEARCLAW/$AstroBEAR$ code may be found online, at http://www.pas.rochester.edu/{\tiny $\sim$}bearclaw/}.  The numerical code solves the conserved form of the equations of fluid dynamics with radiative cooling provided by the cooling curve of \cite{dalgarno1972} and a monotonic ideal gas equation of state. The \emph{AstroBEAR} code has been tested on a variety of problems \citep[see e.g.][]{poludnenko2002,varniere2005,cunningham2006}.

Six runs altogether were undertaken, as discussed below.  The resolution of the simulations were 312 by 156 on the coarsest level with three levels of refinement (for a maximum resolution of 2496 by 1248).  This resulted in 52 and 16 cells per jet and clump radius, respectively, on the finest level. Running on 8--12 processors, each simulation took roughly three days to complete.  All simulations were allowed to run until the jet propagated off the end of the domain, which occurred between t=162--554 years, depending on the simulation.

We have used the observations of \cite{bally2002} as motivation for our investigation.  In our simulations, a jet 100 AU in radius with number density $n_j=10^3\ cm^{-3}$ and temperature $T_j=10^4\ K$ propagates into an ambient region 4800 by 2400 AU in extent having number density $n_a=100\ cm^{-3}$ and temperature $T_a=2,000\ K$. The jet source is varied sinusoidally, 

\begin{equation}
v(t)=v_j + v_a \sin \frac{2\pi t}{\tau}
\end{equation}
where $v_j=150\ km/s$, $v_a=75\ km/s$, and $\tau=30$ years. This velocity modulation creates a series of internal pulsed bowshocks along the jet, as expected \citep{raga1992}.  

We introduce dense circular ``clumps'' into the environment with radii of 30 AU and temperature set so as to maintain pressure balance with the ambient medium. The clouds are all the same size and have a hyperbolic tangent density profile,

\begin{equation}
n(r) = \frac{n_c+n_a}{2}\left(1-\frac{\tanh(r-0.1r_c)}{\tanh(0.1r_c)}\right)\ ,
\end{equation}
where the inclusion of the $0.1r_c$ factor serves to smooth the outer 10\% of the clump, keeping most of the clump area at a constant density $n_c$. The number of clumps and their density are varied in the six runs---see Table 1.

The positions of the clumps were seeded using random number generation with the same seed for each simulation. This allowed for continuity across simulations, as those clumps present for $N_c=60$ exist also for $N_c=120$, and those at $N_c=120$ carry over to $N_c=240$. A selection algorithm was applied so that the minimum allowed separation between clumps would be $2r_c$. The clumps were chosen to fill the restricted domain (x=500--4800 AU, y=0--2400 AU) so that the jet head would have time to form before impacting the condensations.

Passive tracers are used to identify both clump and jet material in the simulation. This proves to be most useful in the analysis of vorticity and material mixing, as discussed in \S~4.

In figures comparing the six runs A--F, we present them in approximate order of increasing clump influence.  The baseline case of a jet propagating in a smooth medium without any clumps is Run A.  Defining the case of 120 clumps with a density ratio between clumps and the jet of $n_c/n_j=100$ as the ``standard'' setup, we investigate the effect of clump density in two runs, with density ratios $n_c/n_j=1$ and $n_c/n_j=10$ denoted as Run B \& C, respectively. The clump baseline run itself is Run E.  Finally, we investigate the effect of varying filling factor by halving and doubling the number of clumps in the domain in Runs D (60 clumps) and F (240 clumps), respectively.

We define the ``spatial filling factor'' as the ratio of the total area of clumps to that of the ambient environment, $f= (N_c \pi r_c^2/A_a)$, where $r_c$ and $N_c$ are clump radius and number, respectively, as given in Table 1, and $A_a$ is the ambient area. We further define the ``dynamic filling factor'' as the clump-to-ambient mass ratio, $f_d=(\rho_c/\rho_a) f$.

\begin{table}\centering
\begin{tabular}[th]{lccccc}
%\multicolumn{6}{c}{{\sc Table 1}}\\
%\multicolumn{6}{c}{{\sc Parameters for the Six Simulations}}\\
\hline\hline
Run & $n_c$ [cm$^{-3}$] & $n_c/n_j$ & N$_c$ & $f$ & $f_d$\\
\hline
% run         nc         c/j   Nc    ff      ffd
A\dotfill   & ---      & --- & --- & ---   & ---   \\
B\dotfill   & 10$^3$   & 1   & 120 & 0.03  & 0.295 \\
C\dotfill   & 10$^4$   & 10  & 120 & 0.03  & 2.95  \\
D\dotfill   & 10$^5$   & 100 & 60  & 0.015 & 14.7  \\
E\dotfill   & 10$^5$   & 100 & 120 & 0.03  & 29.5  \\
F\dotfill   & 10$^5$   & 100 & 240 & 0.06  & 58.9  \\
\hline
\end{tabular}
\caption{Parameters for the six simulations A--F. See text for details.\label{tparam}}
\end{table}

\section{Description of Simulations}
Figure 2 gives the results for the simulations at the same time, $t=180$ years, with the images ordered top-to-bottom as in Table 1.

\begin{figure}[htbp]
\epsscale{0.4}
\plotone{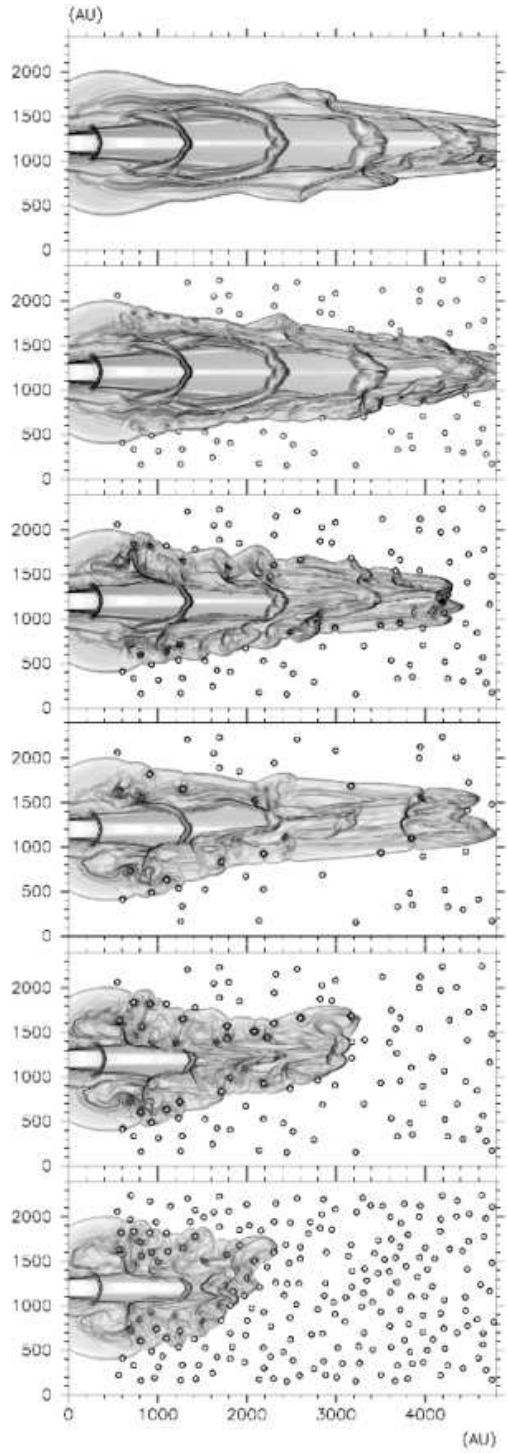}
\caption{Synthetic Schlieren image of Runs A--F all at the same time in the simulation. See text for details.}
%\label{f2}
\end{figure}

\begin{enumerate}
\item The jet-only case (with the jet already propagated off the grid at this time) shows a smooth bowshock with some disruption due to Kelvin-Helmholtz instabilities \citep{blondin1990}. Internal working surfaces are coherent.
\item The next image shows the jet with 120 clumps of density ratio $n_c/n_j=1$. The basic shape of the bowshock is the same as in A), but clumps now ``pockmark'' the surface, causing slight localized disruptions. Internal working surfaces, and the jet's propagation, are essentially unchanged. The conclusion is that clumps at this density ratio are too light to noticeably affect the jet.
\item The third image has an increased density ratio of $n_c/n_j=10$. Here, the bowshock is more clearly disrupted. While the bowshock feature near ($x=2200$ AU, $y=1900$ AU) is still present, other features of the jet's K-H rollovers have been obscured by the clumps. Reverse-facing bowshocks around clumps are present, as is the disruption of the first internal working surface. For the first time, the propagation of the jet has been retarded. The head of the jet is also wider than before, indicative of its disruption.
\item Now the density ratio has increased to its maximum value of $n_c/n_j=100$, but the number of clumps has been halved, from 120 to 60. While the overall shape of the bowshock is similar to the jet-only case of A), the clumps are seen to have great influence in their local neighborhoods. Prominent reverse-facing bowshocks are seen, as is the disruption of the internal working surfaces beyond a certain point (roughly $x=1800$ AU at this time in the simulation). The width of the bowshock near the leading edge is again widened as compared to Run A. However, its propagation is not as slowed as in the previous image, suggesting a greater dependence on filling factor than density ratio---see the next section for a quantitative assessment of this effect.
\item The density ratio remains the same, and the number of clumps is brought back to 120. The effect of the clumps both on propagation and morphology is great. Internal working surfaces are disrupted to the degree that they are ``piling up'' near $x=2000$ AU at this point in the simulation. The clumps have also served to deflect some of the jet's momentum radially (upwards): the feature at ($x=3200$ AU, $y=1700$ AU), engendered by clumps in the range ($x=1800-2400$ AU, $y\sim$1100 AU), will continue to travel at an angle $\sim$30$^{\circ}$ from horizontal. The width near the jet head has increased substantially, and the propagation has been reduced by $\sim$38\% from Run A.
\item Finally, the number of clumps is doubled from 120 to 240. Here, the jet's propagation, disruption, and deflection are all vastly different from the jet-only case of Run A. A large part of the jet is being deflected upwards at an angle of $\sim$25$^{\circ}$, and the jet is dominated by small shocks around the clumps. The propagation of the jet has been reduced significantly, by roughly $58\% $.
\end{enumerate}
As can be seen in the bottom two panels of Fig. 2, the clumps can indeed have a significant impact on deflecting the jet, though it requires high density ratios and numbers.  Whether or not this jet deflection is a transient phenomenon depends highly on the particulars of the clump distribution.  Unless the distribution displays a high degree of coherent structure, subsequent pulses of the jet will eventually bore a hole through the clumps, sweeping them to the side or entraining them. Eventually, the jet will reestablish flow primarily along the axis of the jet. This effect would probably change a great deal in 3D geometry, as discussed in \S~5.

To illustrate the evolution of the jet as it propagates through the inhomogeneous medium, we present Fig. 3. Here, Run E ($n_c/n_j=100$, $N_c=120$) is shown at six times throughout the simulation, every $\sim$90 years from $t=0$ to $440$ years. As the jet pulse has a period of 30 years, the internal working surfaces appear to be at the same location throughout the images, while in truth they capture every third pulse. The panels, from top-to-bottom, show

\begin{figure}[htbp]
\epsscale{0.4}
\plotone{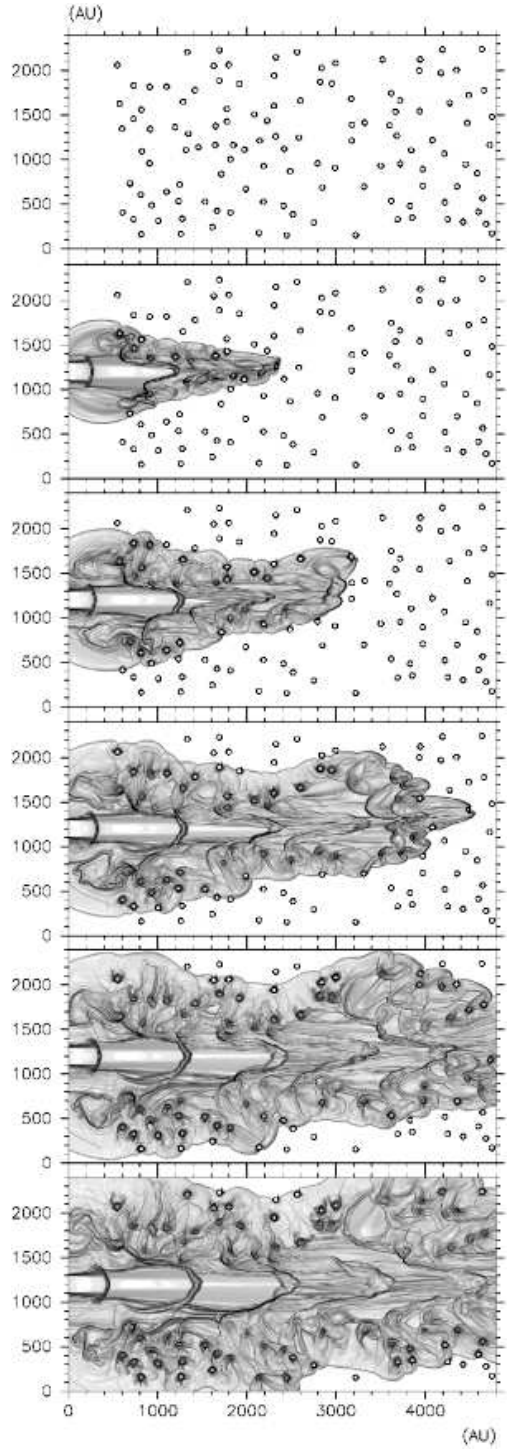}
\caption{Synthetic Schlieren image of Run E at six times throughout the simulation. See text for details.}
%\label{f3}
\end{figure}

\begin{enumerate}
\item The initial clump-filled domain. (Note that a description of the clump-seeding algorithm is given in the preceding section.)
\item The jet has already propagated halfway across the domain. Several clumps near the jet axis are being overrun by the jet's bowshock, though only a few reverse shocks have formed.
\item The jet propagation has slowed, largely due to momentum being redirected at an angle of $\sim30^{\circ}$ from horizontal. A consequence of the deflection is the stagnation of material on the axis, which will be swept away between this and the subsequent image.
\item The head of the jet has largely reclaimed the leading position, though the effects of deflection both above- and below-axis remain evident in the width and angle of bowshocks on the wings. Note also the ``swirling'' motion near the base of the jet, due to the complex interaction between the jet launching and nearby clumps.
\item The leading edge of the jet has propagated off the domain, and the jet's bowshock continues to expand radially. With few exceptions, no clumps which were originally within the jet's radius now remain. Additionally, the interior of the jet cocoon is almost entirely dominated by reverse-facing bowshocks.
\item Finally, in the last frame, most clumps which still exist possess well-developed bowshocks. Many of those near $x=1000$ AU exhibit vertical ``streaming'' features whereas those downstream have broader shapes. All clumps within the jet radius have been swept clear, and as a result the internal working surfaces are beginning to recover coherent structure.
\end{enumerate}

\section{Analysis}
The propagation of the jet through the domain may be investigated by tracking the leading edge of the jet throughout the simulation, as shown in Fig. 4.  Here again the effect of the clumps on the jet is clear, as is the fact that the number of clumps has a far greater effect than the density thereof.  The crossing times in years for Runs A--F are, respectively, 
\begin{equation}
t_{end}=(162, 165, 202, 180, 277, 554)\ .
%\label{t_end}
\end{equation} 
In the jet-only case, this corresponds to a bowshock speed of $\sim$30 AU/year. This may be compared to the theoretical bowshock speed of \cite{blondin1990},

\begin{equation}
v_{bs}\simeq\frac{v_j}{1-\sqrt{n_a/n_j}}
\end{equation}
which for $v_j=150\ km/s$ gives a speed of $v_{bs}\simeq32$ AU/year and a crossing time of 150 years, in fair agreement.

The variability of the speed of the leading edge between simulations suggests two natural perspectives with which to discuss their evolution---namely, time and the position of the leading edge of the jet. Comparisons involving the former compares processes occurring at the same absolute time in the simulation, the latter serves to effectively describe them on the same scale of jet propagation through the medium. While both are legitimate, the latter perspective lends itself to a more illuminating investigation of the simulation suite in nearly all details. The investigation of vorticity benefits from both descriptions, as will be seen.

\begin{figure}[htbp]
\epsscale{0.5}
\plotone{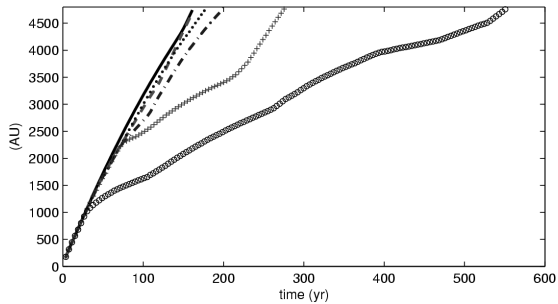}
\caption{The leading edge of the jet material as a function of time for each run. Symbols for each run are Run A: solid, Run B: dashed, Run C: dot-dashed, Run D: dotted, Run E: $+$, and Run F: $\circ$.}
%label{f4}
\end{figure}

Using a nonlinear least-squares fit, the time needed for the jet to reach the edge of the grid is relatively sensitive to the clump number ($t_{end}\sim N_c^{1.5}$), and is only weakly dependent on their density ($t_{end}\sim n_c^{0.31}$). Furthermore, the dependence of grid-crossing time on the factor $f_d$ is approximately squared ($t_{end}\sim f_d^{2.05}$); see Fig.~5       .

\begin{figure}[htbp]
\epsscale{0.5}
\plotone{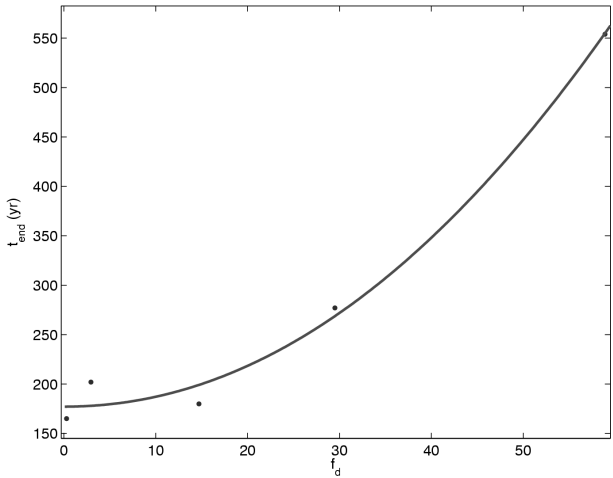}
\caption{A power law fit to the crossing time $t_{end}$ as a function of the dynamic filling factor $f_d$ for Runs B-F. The power law fit corresponds to an index of 2.05.}
%label{f5}
\end{figure}

With such a complex interaction between jet and clumps, the questions of mass loading and momentum entrainment come to the fore. Namely, is all the clump material swept to the sides by the jet's bowshock or is there significant bulk axial momentum imparted to the clumps?  How much and in what manner does the jet and clump material mix?

\begin{figure}[htbp]
\epsscale{0.5}
\plotone{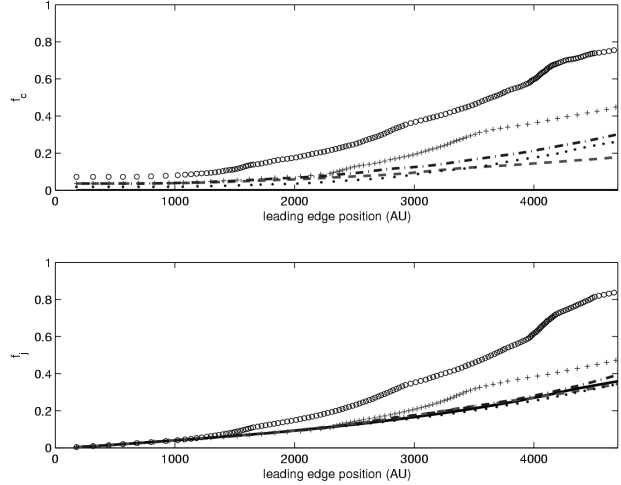}
\caption{The evolution of the clump filling factor $f_c$ and jet filling factor $f_j$ as functions of the position of the leading edge of the jet. The top panel is $a$, the bottom $b$. Line styles are as in Fig. 4.}
%label{f6}
\end{figure}

By introducing the clump filling factor $f_c^R$, similar to $f$ but for a given run $R$, and the jet filling factor $f_j^R$ (analogously), we may investigate the evolution of their respective mixing and diffusion as the jet propagates through the grid. Figure 6$a$ \& $b$ show $f_c^R$ and $f_j^R$, respectively, versus the position of the jet leading edge. The mixing is seen to be coupled with the dynamic filling factor $f_d$. By the time the jet reaches the end of the domain, the $f_c$ value for Runs B--D is only $\sim$0.25, but for E and F is $0.47$ and $0.77$, respectively. This stands to reason as the slower propagation speeds in Runs E and F allow more time for the clumps behind the relatively stalled jet head to be disrupted into the environment. Similarly, Panel $b$ demonstrates that the jet's lateral expansion is enhanced by increased $f_d$. By the time the jet has reached the edge of the grid, the jet filling factor for Run F is more than twice that of Run A ($f_j^A=0.35$ and $f_j^F=0.84$). The measure of $f_d$ thus provides a quantitative description of the disruption of the jet by the clumps: though its forward progress is slowed, ultimately the jet will increasingly fragment and fill the domain as it progresses through the medium.

\begin{figure}[htbp]
\epsscale{0.5}
\plotone{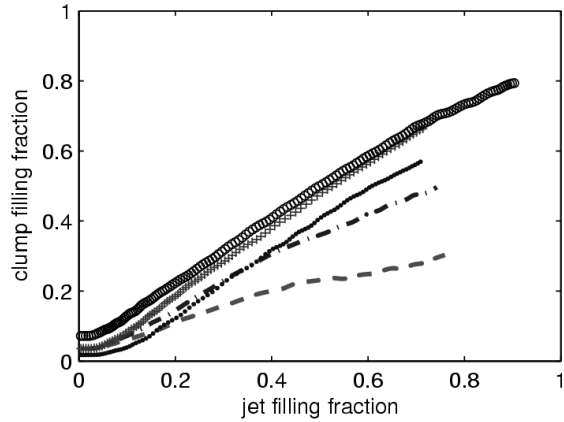}
\caption{Clump filling factor $f_c$ plotted as a function of jet filling factor $f_j$ for the five runs B--F. Line styles are as in Fig.~4.}
%\label{f7}
\end{figure}

The fact that the jet and clump filling factors depend similarly on $f_d$ suggests that there may be a correspondence between $f_d$ and the coupling between jet and clump material. Figure~7 demonstrates this: there exists a higher correlation---defined by the slope of the linear trends of the data---between $f_c$ and $f_j$ for higher $f_d$. As $f_d$ increases, the correlation number relating the two quantities rise from $\sim$0.35 for Run B to $\sim$0.80 for Runs D--F. This is to be expected for as jet material is deflected by the clumps, clump material is spread by the jet.  The mixing of both materials are indeed dynamically coupled.

\begin{figure}[htbp]
\epsscale{0.5}
\plotone{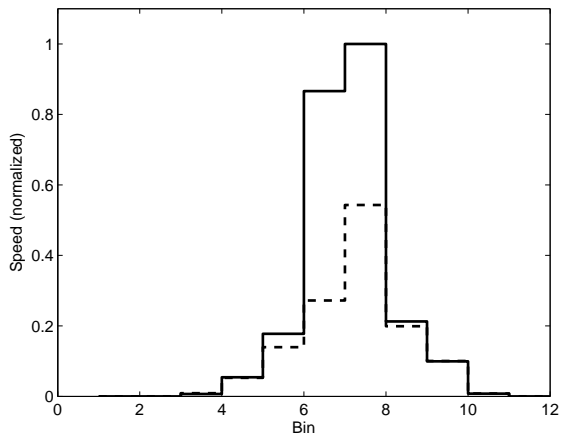}
\caption{Binned and normalized speed of jet and clump material for Run E at the same time as in Fig.~2. Domain is binned into 12 sections along y, with velocity summed over the entire x-extent for each bin, then normalized over the collection. Jet material is solid line and clump material dashed.}
%\label{f8}
\end{figure}

Momentum entrainment also appears to occur, as shown in Fig.~8.  In this figure we divide the radial extent (i.e., along the $y$-axis) into twelve bins.  Within each bin we sum over the entire axial extent of the grid (i.e., over all $x$) the axial component of the velocity, and normalize the results. Then, the bin with the highest velocity may be compared with bins which show less motion.  Here we see that the further out one progresses radially, the less the axial momentum is transferred to the clumps.  These data may be compared with Fig. 17 of \cite{bally2002}, which analogously gives the proper motion of features as a function of distance from the flow axis. The figures are in fair agreement.

\begin{figure}[htbp]
\epsscale{0.5}
\plotone{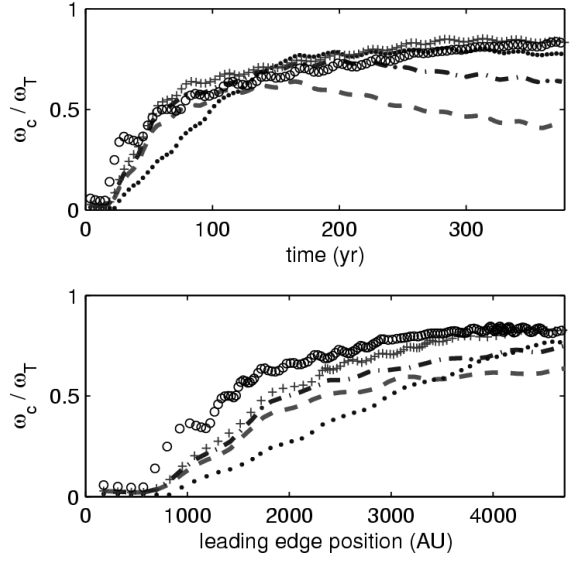}
\caption{The fractional clump vorticity $\omega_c/\omega_T$ as a function of time and position of the leading edge of the jet for runs B--F. Line styles are as in Fig.~4.}
%\label{f9}
\end{figure}

\begin{figure}[htbp]
\epsscale{0.5}
\plotone{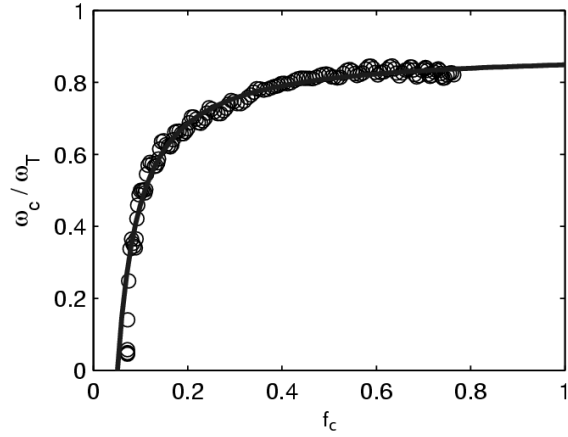}
\caption{A power law fit to the fractional clump vorticity $\omega_c/\omega_T$ as a function of the clump filling factor $f_c$ for Run F. The power law fit corresponds to an index of -1.1.}
%\label{f10}
\end{figure}

The possibilities of turbulent clump mixing is addressed by Fig.~9. The total magnitude of the vorticity of all material in the grid is dominated by processes associated with the jet flow (e.g., the K-H rollover). In lieu of investigating this total vorticity $\omega_T$, we investigate the evolution of the fractional clump vorticity $\omega_c/\omega_T$ as determined by the clump tracer versus time and leading edge position in Fig.~9. The top panel, in conjunction with the crossing-times listed in Eq.~3, demonstrates that the fractional vorticity flattens (Runs D \& E) or decreases (Runs B \& C) once the jet propagates off the grid. Although not depicted on the plot, Run F eventually flattens to a value near that of Run E. Further, the rate of falloff varies inversely with $f_d$, suggesting that the ability of an environment to sustain vorticity is directly correlated to the clump densities within that environment. 

In the bottom panel, two effects are seen. The first is that increasing clump density $n_c$ affects the scale of fractional vorticity in a regular manner: the increase in final $\omega_c/\omega_T$ from Run B to C and Run C to E are the same, showing a consistent increase of 11\% between each pair. Secondly, an increased clump number $N_c$ affects the concavity of the fractional vorticity's evolution. That is, while the final values are nearly the same for Runs D-F, the rate at which this value is attained increases markedly with number $N_c$.

Additionally, the dependence of fractional clump vorticity $\omega_c/\omega_T$ on clump filling factor $f_c$ has a power law index of roughly -1.1. This relation is shown in Fig.~10 for Run F. However, it holds for all Runs B--F, indicating that the generation of turbulent motion and clump dispersal are well correlated.

\begin{figure}[htbp]
\epsscale{0.5}
\plotone{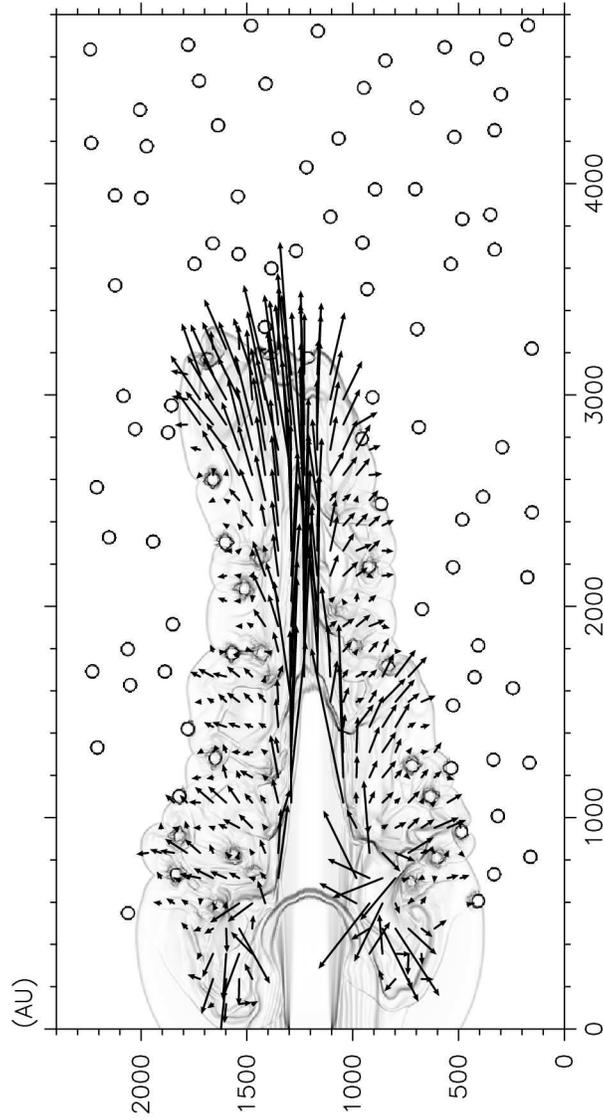}
\caption{A synthetic Schlieren image of Run E at the same time as in Fig.~2, now with velocity vectors overlaid. Details of note include large off-axis vectors near the leading edge and reverse-pointing vectors near the base of the jet due to vorticity. A superficial correspondence may be seen between this and Fig.~1.}
%\label{f11}
\end{figure}

Finally, as a purely qualitative result we present Fig.~11, which is a snapshot of the standard run with an array of velocity vectors overplotted for regions where the clump tracer and velocity are nonzero. This may be compared with Fig.~1 and some basic similarities can be seen.  Note that dispersal of velocity vectors near the head showing the change in direction of the jet.  The redirection of the flow near clumps is also apparent.  Additionally, vectors further from the jet axis are roughly half or less the magnitude of those near the axis as discussed previously. Of further note are the velocity vectors near the jet launching region which are directed perpendicularly or even backwards, indicative of the vortical motions present within that region.

\section{Discussion and Conclusion}
We have presented simulations of a pulsed astrophysical jet propagating into a heterogeneous medium which possesses small-scale condensations.  We find that the condensations can have an effect on the jet propagation which is strongly correlated to the filling factor of the distribution and less so on the density contrast between the jet and clumps.

It is reasonable to anticipate that the slowing of the jet will allow increased time to mix clumps, resulting in increased filling factors for the clumps. If the clumps and the jet are dynamically coupled, this will result in an increased jet filling factor as well. Indeed, when considered against the position of the leading edge, both quantities are noticeably enhanced with increased $f_d$.

Similarly, the fraction of the vorticity in the clumps has a clear dependence on the dynamic filling factor when considered against the position of the jet's leading edge: increasing $f_d$ results both in a higher fraction of the vorticity attributed to the clumps and in the earlier realization of this higher fraction. It may therefore be said that $f_d$ is a robust indicator of the generation and sustainability of vorticity in the domain.

Qualitatively, our results can be considered in the broader context of astrophysical jet research.  In particular, the morphologies presented here bear resemblance to those of \cite{saxton2005}, who performed 2D relativistic hydrodynamic simulations of overpressured, underdense AGN jets propagating into a heterogeneous medium. Unlike the present work, the heterogeneity was seeded via a Fourier algorithm intended to mimic power-law structure formation by turbulence, resulting in a variety of shapes and sizes of inhomogeneities. Despite this disparity in the scenarios, both simulations shared many morphological features, perhaps suggestive of scale-robust physics. See for example Figure 6 of that work, in which a jet is deflected off-axis in much the same manner as our standard run. If we cast those runs in terms of the dynamic filling factor $f_d$ as defined here, their simulations take on values 4.11--118.0, covering some of the same range as in the present study. While one might be tempted to assume the highest correlation between the two studies would occur for runs with the most-similar $f_d$, the variable sizes of inhomogeneities in the previous work prevent such a strict correspondence to be drawn. In fact, the two simulations most similar are that study's Run A1 and the present Run E, with values for $f_d$ of 14.6 and 29.5, respectively.

Both of the results given in Figures~7 and 8 suggest that viable mass loading and momentum entrainment can occur.  This is in contrast to the mechanism discussed by \cite{raga2003}. In that case, an object impinged upon the side of a steady propagating jet and was found to entrain only transiently. One possible explanation (as mentioned in that work) is that whereas the contribution from a single clump would have little long-term effect, the continued entrainment by successively over-run clumps serves to maintain the mass loading. 

Another result from our study is the realization that objects on size scales $r_c < r_j$ can have a large influence on an astrophysical jets.  Our simulations show clumps can offer a significant disruption mechanism in radiative (stellar) jets which has not before been considered.

As noted however in the preceding section, many of the results presented here would doubtless be different in a 3D geometry.  In particular, one could imagine that the impediment presented by the clumps would be diminished due to the additional degree of freedom. This suggests that the recovery of similar results would require either higher density ratios or larger filling factors. 

Additionally, magnetic fields appear to have an effect on jet propagation \citep[see e.g.][]{frank1998, gardiner2000, gardiner2001, decolle2005}, and the fact that magnetic confinement has been advanced as a possibility for maintaining small-scale structure \citep{diamond1989} suggests MHD simulations in this parameter regime may be useful.

We now consider the observational evidence for structures on these scales beginning for reference with the ISM and then examining molecular clouds.  As we shall see the issue remains contentious. The presence of structure on 10--100 AU scale in either the ISM or star-forming regions is not yet settled. 

\cite{heiles1997} demonstrated the difficulty of having dense ($10^4-10^5\ cm^{-3}$), small-scale ($\sim$15 AU) atomic structure be in pressure balance with the ISM due to the necessarily low temperatures involved. Of the three cooling mechanisms mentioned---the absence of dust grains, cooling via CO refrigeration, and an attenuated interstellar radiation field---only CO cooling was considered reasonable to entertain, though it was still unsatisfying. As an alternative, he put forth the idea that apparent small-scale features are in fact edge-viewed larger structures such as filaments or sheets. Heiles demonstrated that the requirement on pressure balancing is ameliorated somewhat by this concept, and discussed the observational consequences. (However, this conclusion does not directly negatively affect our results, as a 2D slab-symmetric simulation mimics exactly these types of structures.)

Subsequently, \cite{deshpande2000} presented analytic results purporting that small-scale structure in the ISM does not need to exist. Instead, irregularities in large-scale structure can mimic small-scale structure. This conclusion was reached by extending the power-law spectrum of structure on scales $\sim$0.1 pc down to 10 AU. 

However, \cite{faison2001} presented new observations which seemed to contest this assertion. Using very long baseline interferometry, Faison and Goss were able to resolve absorbing features of order 20 milliarcseconds. Using 21cm absorption of galactic H I toward several bright extragalactic continuum sources, they were able to resolve several distinct objects with inferred densities $\sim$10$^4\ cm^{-3}$ and scales $\sim$10--100 AU. Of the four sources imaged, three returned results which appeared in line with Deshpande's treatment (a variation in optical depth of about 0.1). One, however, showed a change in opacity of 0.6--0.8, quite in excess of the prediction. Moreover, Deshpande's analysis suggested that only gradients in optical depth should be observed since the variations were coming from larger-scale structure. However, in each case presented by Faison and Goss, distinct objects were apparent, in direct contradiction to that prediction.

Relatedly, \cite{lauroesch2000} have investigated small-scale structure via Na I absorption toward a binary star system. While small-scale structure did seem present, the densities were significantly lower than those typically claimed (only $\sim$20-200 $cm^{-3}$).

For star-forming regions, the presence and prevalence of structure on this scale is less clear. For the densities and temperatures used here, emission would be expected either in molecular lines (for the low-contrast clumps) or lines more reminiscent of cloud cores (for the heavier clumps) \citep{vannier2001}. In an investigation of structure sizes in the Orion molecular cloud, \cite{vannier2001} found a preferential size scale between $3\cdot10^{-3}$ and $4\cdot10^{-3}$ pc ($600$ and $800$ AU) via area-perimeter, Fourier, and brightness distribution analysis of H$_2$. However, their study was biased against small-scale structure due to a limiting resolution of $0.15$'', corresponding to a physical length of $50$ AU. In particular, the model against which the Fourier data was vetted showed spurious small-scale signals due to noise below $\sim$5$\cdot10^{-4}$ pc (100 AU) with a maximum at $3.5\cdot10^{-4}$ pc (75 AU). As this peak was also featured in the real data, these and smaller scales were not included. Attempting to directly measure and count structures was once again impacted by the resolution of 0.15''. These results suggest that structure below these scales in Orion is at least ambiguous, if not wholly endorsed.

As to prevalence, \cite{hatchell2005} imaged the complete Perseus molecular cloud at 850 and 450 $\mu$m in order to map locations and conditions for star formation. However, their study surveyed cloud cores only down to a mass of 0.4$M_\odot$. Assuming spherical symmetry, a single clump in the present work is substantially smaller than this limit. The \cite{falgarone2004} review noted that, considering the positive correlation between gas mass and size scale, and assuming that this trend holds for all scales, then it is reasonable to assume that small-scale structure cannot comprise a notable fraction of the molecular cloud mass.  This does not however preclude their presence  in regions close to cloud cores where stars have recently formed and where jet propagation is initiated.

It should be noted that pressure balance between the clumps and ambient medium forces the clump temperatures to the range 2--200 K. Similar overall parameters have been simulated in prior studies; in many cases, pressure balance appears to have been sacrificed to avoid egregiously cold clumps \citep[e.g.][]{raga2003,melioli2005}. While the simulations are most likely robust to the exact details of the temperatures, this aspect of the parameter regime deserves further careful consideration.

Our conclusion is that, though the presence of small clumps remains uncertain, our results address significant dynamics which have heretofore been unexplored.  Jets clearly show clumpy morphology and we have explored one means by which these morphologies can be achieved and the consequences of propagation in heterogeneous media.  We can also suggest a new direction in which to turn in consideration of clumpy jets. While clumps may form from instabilities associated with cooling at the jet head, its is also possible the jet beams are themselves clumpy.  Future studies will explore distinct clumps along the axis in the beam which interact internally with the jet. It is reasonable to suppose that a clump so launched could overtake a bowshock within the jet and disrupt it. The exact details of this interaction should be a matter of some interest.

\acknowledgements
The authors acknowledge support for this work by the $Spitzer$ Space Telescope theory grant 051080-001, National Science Foundation grant AST 05-07519, and Department of Energy grant DE-F03-02NA00057. We also acknowledge the computational resources provided by the Laboratory for Laser Energetics at the University of Rochester.


\begin{thebibliography}


\bibitem[Bally et al.(2002)]{bally2002} Bally, J., Heathcote, S., Reipurth, B., Morse, J., Hartigan, P., and Schwartz, R. 2002, AJ, 123, 2627
\bibitem[Blondin, Fryxell, \& K\"onigl(1990)]{blondin1990} Blondin, J.M., Fryxell, B.A., and K\"onigl, A. 1990, ApJ, 360, 370
\bibitem[Cabrit \& Raga(2000)]{cabrit2000} Cabrit, S. and Raga, A. 2000, A\&A, 354, 667
\bibitem[Cunningham, Frank, \& Blackman(2006)]{cunningham2006} Cunningham, A.J., Frank, A., and Blackman, E.G. 2006, ApJ, 646, 1059
\bibitem[Dalgarno \& McCray(1972)]{dalgarno1972} Dalgarno, A. and McCray, R. 1972 ARA\&A, 10, 375
\bibitem[de Gouveia Dal Pino(1999)]{dalpino1999} de Gouveia Dal Pino, E.M. 1999, ApJ, 526, 862
\bibitem[De Colle \& Raga(2005)]{decolle2005} De Colle, F. and Raga, A.C. 2005, MNRAS, 359, 164
\bibitem[Deshpande(2000)]{deshpande2000} Deshpande, A.A. 2000, MNRAS, 317, 199
\bibitem[Diamond et al.(1989)]{diamond1989} Diamond, P.J., Goss, W.M., Romney, J.D., Booth, R.S., Kalberla, P.M.W., and Mebold, U. 1989, ApJ, 347, 302
\bibitem[Faison \& Goss(2001)]{faison2001} Faison, M.D. and Goss, W.M. 2001, AJ, 121, 2706
\bibitem[Galgarone, Hily-Blant, \& Levrier(2004)]{falgarone2004} Falgarone, E., Hily-Blant, P., and Levrier, F. 2004, Ap\&SS, 292, 89
\bibitem[Fragile et al.(2004)]{fragile2004} Fragile, P.C., Murray, S.D., Anninos, P., and van Breugel, W. 2004, ApJ, 604, 74
\bibitem[Frank et al.(1998)]{frank1998} Frank, A., Ryu, D., Jones, T.W., and Noriega-Crespo, A. 1998, ApJ, 494, 79
\bibitem[Gardiner et al.(2000)]{gardiner2000} Gardiner, T.A., Frank, A., Jones, T.W., and Ryu, D. 2000, ApJ, 530, 834
\bibitem[Gardiner \& Frank(2001)]{gardiner2001} Gardiner, T.A., and Frank, A. 2001, ApJ, 557, 250
\bibitem[Hatchell et al.(2005)]{hatchell2005} Hatchell, J., Richer, J.S., Fuller, G.A., Qualtrough, C.J., Ladd, E.F., and Chandler, C.J. 2005, A\&A, 440, 151
\bibitem[Hartigan et al.(2005)]{hartigan2005} Hartigan, P., Heathcote, S., Morse, J.A., Reipurth, B., and Bally, J. 2005, AJ, 130, 2197
\bibitem[Heiles(1997)]{heiles1997} Heiles, C. 1997, ApJ, 481, 193
\bibitem[Hennebelle \& Passot(2006)]{hennebelle2006} Hennebelle, P. and Passot, T. 2006, A\&A, 448, 1083
\bibitem[Lauroesch, Meyer, \& Blades(2000)]{lauroesch2000} Lauroesch, J.T., Meyer, D.M., and Blades, J.C. 2000, ApJ, 543, L43
\bibitem[Masciadri et al.(2002)]{masciadri2002} Masciadri, E., de Gouveia Dal Pino, E.M., Raga, A.C., and Noriega-Crespo, A. 2002, ApJ, 580, 950
\bibitem[Melioli, de Gouveia Dal Pino, \& Raga(2005)]{melioli2005} Melioli, C., de Gouveia Dal Pino, E.M., and Raga, A. 2005, A\&A, 443, 495
\bibitem[O'Sullivan \& Lery(2002)]{osullivan2002} O'Sullivan, S. and Lery, T. 2002, RevMexAA, 13, 98
\bibitem[Poludnenko, Frank, \& Blackman(2002)]{poludnenko2002} Poludnenko, A.Y., Frank, A., and Blackman, E.G. 2002, ApJ, 576, 832
\bibitem[Raga \& B\"ohm(1987)]{raga1987} Raga, A.C. and B\"ohm, K.H. 1987, ApJ, 323, 193
\bibitem[Raga \& Kofman(1992)]{raga1992} Raga, A.C. and Kofman, L. 1992, ApJ, 386, 222
\bibitem[Raga et al.(2002)]{raga2002} Raga, A.C., de Gouveia Dal Pino, E.M., Noriega-Crespo, A., Mininni, P.D., and Vel\'azquez, P.F. 2002, A\&A, 392, 267
\bibitem[Raga et al.(2003)]{raga2003} Raga, A.C., Vel\'azquez, P.F., de Gouveia Dal Pino, E.M., Noriega-Crespo, A., and Mininni, P. 2003, RevMexAA, 15, 115
\bibitem[Reipurth, Bally, \& Devine(1997)]{reipurth1997} Reipurth, B., Bally, J., and Devine, D. 1997 AJ, 114, 2708
\bibitem[Reipurth \& Bally(2001)]{reipurth2001} Reipurth, B. and Bally, J. 2001 ARA\&A, 39, 403
\bibitem[Rosen \& Smith(2004)]{rosen2004} Rosen, A. and Smith, M.D. 2004, A\&A, 413, 593
\bibitem[Saxton et al.(2005)]{saxton2005} Saxton, C.J., Bicknell, G.V., Sutherland, R.S., Midgley, S. 2005, MNRAS, 359, 781
\bibitem[Steffen \& L\'opez(2004)]{steffen2004} Steffen, W. and L\'opez, J.A. 2004, ApJ, 612, 319
\bibitem[Varniere et al.(2005)]{varniere2005} Varniere, P., Poludnenko, A., Cunningham, A., Frank, A., and Mitran, S. 2005 in {\emph Adaptive Mesh Refinement---Theory and Applications}, ed. T. Plewa, T. Linde, \& G.V. Weirs (Berlin: Springer) 453
\bibitem[Vannier et al.(2001)]{vannier2001} Vannier, L., Lemaire, J.L., Field, D., Pineau des For\^ets, Pijpers, F.P., and Rouan, D. 2001, A\&A, 366, 651
\bibitem[V\"olker et al.(1999)]{volker1999} V\"olker, R., Smith, M.D., Suttner, G., and Yorke, H.W. 1999, A\&A, 343, 953


\end{thebibliography}
\end{document}